\documentclass[twocolumn,showpacs,preprintnumbers,amsmath,amssymb]{revtex4}

\usepackage[dvips]{graphicx}
\usepackage{dcolumn}
\usepackage{bm}
\usepackage{amssymb}
\usepackage{array}
\usepackage{longtable}

\begin{document}

\preprint{APS/123-QED}

\title{Traffic fluctuations on weighted networks}

\author{Yichao Zhang$^{1,2}$}

\author{Shi Zhou$^{2}$}
\email{s.zhou@cs.ucl.ac.uk}

\author{Zhongzhi Zhang$^{3,4}$}
\email{zhangzz@fudan.edu.cn}

\author{Jihong Guan$^{1}$}
\email{jhguan@tongji.edu.cn}

\author{Shuigeng Zhou$^{3,4}$}
\email{sgzhou@fudan.edu.cn}

\author{Guanrong Chen$^{5}$}
\email{gchen@ee.cityu.edu.hk}

\affiliation{$^{1}$Department of Computer Science and Technology,
Tongji University, 4800 Cao'an Road, Shanghai 201804, China}

\affiliation{$^{2}$Department of Computer Science, University
College London, Gower Street, London, WC1E 6BT, United Kingdom}

\affiliation {$^{3}$School of Computer Science, Fudan University,
Shanghai 200433, China}

\affiliation {$^{4}$Shanghai Key Lab of Intelligent Information
Processing, Fudan University, Shanghai 200433, China}

\affiliation {$^{5}$Department of Electronic Engineering City
University of Hong Kong, Hong Kong, China}

\date{\today}

\begin{abstract}
Traffic fluctuation has so far been studied on unweighted networks.
However many real traffic systems are better represented as weighted
networks, where nodes and links are assigned a weight value
representing their physical properties such as capacity and delay.
Here we introduce a general random diffusion (GRD) model to
investigate the traffic fluctuation in weighted networks, where a
random walk's choice of route is affected not only by the number of
links a node has, but also by the weight of individual links. We
obtain analytical solutions that characterise the relation between
the average traffic and the fluctuation through nodes and links. Our
analysis is supported by the results of numerical simulations. We
observe that the value ranges of the average traffic and the
fluctuation, through nodes or links, increase dramatically with the
level of heterogeneity in link weight. This highlights the key role
that link weight plays in traffic fluctuation and the necessity to
study traffic fluctuation on weighted networks.
\end{abstract}


\maketitle

\section{\label{sec:level1}INTRODUCTION\protect}

In nature and society, many complex systems can be represented as
graphs or networks, where nodes represent the elementary units of a
system and links stand for the interactions between the nodes.
Complex networks have been a research focus in the last decade
~\cite{PRL863200,PRE66065102,PRL91148701,PRE69036102,PRE71026125,PRE65026139,RMP7447,AP511079,SIAMR45167,PR424175}.

Recently attention has been given to the traffic fluctuation problem
in networks. It is associated with an additive quantity representing
the volume of traffic travelling through a node (or a link) in a
time interval, and the dependence between its mean and standard
deviation~\cite{Nature189732}. Knowledge on traffic fluctuation is
relevant to the design and engineering of real systems such as air
transport network, highway network, power-grid network and the
Internet, for example how to deploy network resources, how to route
traffic efficiently and how to mitigate congestion.

In recent years there has been a strong research interest in the
traffic fluctuation problem, which is relevant to a wide range of
applications in various networked
systems~\cite{PRL92028701,PRL93068701,PRL96218702,PRL100208701,IJBC172363,AP5789}.
In particular researchers are interested in the relation between the
mean of traffic $\langle f\rangle$ and the standard deviation
$\sigma$ at a given node. This is because various problems of
immediate social and economical interests are ultimately constrained
by the extent to which the assignment of resources matches supply
and demand under realistic conditions, and the resource assignment
is essentially governed by the `normal' traffic behaviour
characterised by large fluctuations.

In many real systems, traffic fluctuations are often affected by
specific physical properties of network elements, such as the
bandwidth of a cable or the computational power of an Internet
router. Such systems are much better described as a more
sophisticate form of network graphs, the weighted networks, where
the physical properties of network elements are represented by
link's weight and node's strength.

In this paper we investigate the traffic fluctuation problem in
weighted networks. In Section II, we review the previous works on
traffic fluctuations in unweighted networks. In Section III, we
introduce some network properties related to our work and define a
number of variables that are used in the study of traffic
fluctuation. We introduce a general random diffusion (GRD) model,
where a general random walker's choice of path is affected by link's
weight. In Section IV and V, we analyse the fluctuation of traffic
in weighted networks. We provide analytical solutions on the
relation between the fluctuation and the average traffic at nodes in
section IV and on links in section V. We also run numerical
simulations, which confirm our analysis and illustrate its physical
meaning. We summarize our work in Section VI.

Our contributions are four folders.
Firstly, we introduce a more general analytical law which
characterises the traffic fluctuation on weighted networks. Previous
works are a special case of our law.
Secondly, our results show that traffic fluctuations on a weighted
network can be dramatically different from that on the equivalent
unweighted network. This highlights the necessity of studying real
systems as weighted networks when network elements have a non
trivial impact on traffic dynamics.
Thirdly, in addition to traffic fluctuation through nodes, we also
analytically study traffic fluctuations through links. We show that on weighted networks the
traffic fluctuation on links is significant and should be considered
when designing real systems.
Finally, we reveal the dependence between a link's traffic
properties and the connectivity of the link's two end nodes.


\section{Previous Studies on Traffic Fluctuations}\label{PSTF}

The early discovery was that the average volume of traffic arriving
at a node, $\langle f\rangle$, and the fluctuation (standard
deviation) of the traffic, $\sigma$, follow a power-law relation,
i.e.~$\sigma\sim \langle f\rangle^{\alpha}$, where the exponent
$\alpha$ has two universal values, 1/2 and
1~\cite{PRL92028701,PRL93068701}. This result has attracted a lot of interest from the network
research community and it also generated debates. Subsequently, it
has been shown numerically that there is a wide spectrum of possible
values within the range of $[1/2,1]$ for $\alpha$~\cite{PRL96218702}.

Next, Kujawski et al~\cite{NJP9154} revealed some scaling properties of traffic fluctuation by conducting simulations on unweighted scale-free networks, where a navigation algorithm is used to give a preference for less used edges in the traffic history. However the scenario considered in~\cite{NJP9154} is unrealistic. because if an edge is less used in a network's traffic history, it indicates that the edge is indeed not preferred, for reasons like smaller capacity, longer delay or more expensive.

Recently, Meloni et al~\cite{PRL100208701} derive an analytical law
showing that the dependence of fluctuations with the mean traffic on
unweighted networks. They point out the dependence of fluctuations
with the mean traffic is governed by the delicate interplay of three
factors: the size of observation window; the noise associated to the
fluctuations in the number of packets from time window to time
window; the degree of the node. However, unweighted networks are
relatively simple and widely used to represent the connectivity
structure of a network system. On an unweighted network, physical
properties of links (and nodes) are removed such that all links are
equal, i.e.~each link only represents the existence of a topological
connection between two nodes.

As is known, many real systems display different interaction
strengths between nodes, which reveal unweighted networks' drawback
in link definition. In this case, it is easy to realize that traffic
path is rarely randomly chosen. This is because links have different
physical properties (bandwidth, delay or cost) and naturally traffic
tends to choose a path to achieve better performance, higher
efficiency or less cost.

\section{Traffic Fluctuation on Weighted Networks}

\subsection{Weighted Networks}

A more realistic form of networks is the weighted
networks~\cite{PRE70056131,PRE70066149,PNAS1013747}, where each link
is assigned a {\emph weight} value to denote a physical property of
interest, e.g.~the bandwidth of a cable or the length of a road; and
similarly, each node is assigned a {\emph strength}, for example, to
represent the computational capacity of an Internet router. Weighted
networks encode more information and they are a more realistic
representation of real systems where individual links (and nodes)
are vastly different.

Weighted networks have the advantage to encode information of
physical properties of links and nodes. For example in a weighted
social network, a link can indicate that two people know each other
while the weight of the link can denote how often they meet each
other~\cite{PNAS1013747}; in a weighted Internet router network,
link weight can represent the bandwidth of a cable and node strength
can represent the process power of a router~\cite{NJP10053022}; in a
weighted aviation network, link weight can denote the annual volume
of passengers travelling between two airports~\cite{PNAS1013747};
and in the weighed metabolic network \emph{E.~coli}, link weight can
encode the optimal metabolic fluxes between two
metabolites~\cite{Nature427839}. On the other hand, recently there
are some works on random walk based on weighted networks, but they
only considered a single random
walker~\cite{PRE77066105,PRE80016107}.

In this work we study the traffic fluctuation problem in weighted
networks and investigate critical questions such as `what are the
impact of different capacity of nodes and links on the fluctuation
of traffic passing through them?' `can we predict the fluctuation?'
and `what are the implications for network resource assignment?'

\subsubsection{Link Weight}

In network research the degree, $k$, is defined as the number of
links a node has. When representing real systems as weighted
networks, the weight of a link is often related to degrees of the
two end nodes of the link. For example the number of scheduled
flights between two airports increases with the number of flights
each of the two airports has.

In this case, we define the weight of a link between nodes $i$ and
$j$ as
\begin{equation}
w_{ij}=w_{ji}=(k_ik_j)^\theta,\label{wij}
\end{equation}
where $k_i$ and $k_j$ are degrees of the two nodes, and $\theta$ is
the network's weightiness parameter which characterises the
dependence between link weight and the node
degrees~\cite{PNAS1013747,Nature427839,PRL94188702}. This definition
is well supported by empirical
studies~\cite{PNAS1013747,Nature427839,PRL94188702} and is widely
used in researches on weighted networks. It introduces the
weightiness parameter which conveniently determines the level of
link heterogeneity in a weighted network. When $\theta=0$ there is
no dependence between link weight and node degree, all links are
equal with $w=1$, and the network becomes an unweighted network.
When $\theta>0$, it is a weighted network where links have different
weights. The larger $\theta$, and the wider difference between
links.

\subsubsection{Node Strength}

On the other hand, the strength of node $i$ is defined as
\begin{equation}
s_i=\sum_{j\in\Gamma(i)}w_{ij}=\sum_{j\in\Gamma(i)}{(k_ik_j)^{\theta}},
\label{si}
\end{equation}
where $\Gamma(i)$ is the set of neighbours of node $i$. In an
unweighted network with $\theta=0$, $s_i=k_i$ node strength is the
same as node degree. In a weighted network with $\theta>0$, a node's
strength is the sum of the weight of the links connecting to the
node. Two nodes with the same degree may have different strength
values depending on the weight of their links. For example consider
two airports $A$ and $B$, both have 4 flight connections,
$k_A=k_B=4$. Airport $A$ will have more `strength' than airport $B$
if the former is connected with four well-connected hub airports and
the later is connected with four less-connected local airports.

\subsection{General Random Diffusion (GRD) Model}

Random walk is a mathematical formalisation of a trajectory that
consists of taking successive random steps. A familiar example is
the random walk phenomenon in a liquid or gas, known as Brownian
motion~\cite{AP17549,AP21756}. Random walk is also a fundamental
dynamic process on complex networks~\cite{PRL92118701}. Random walk
in networks has many practical applications, such as navigation and
search of information on the World Wide Web and routing on the
Internet~\cite{PRE63041108,PRE64046135,PRL89248701,PRE74046118,PA385743}.
Previous research on traffic fluctuation either studied random
walkers travelling on unweighted networks where the choice of route
is random as all links are regarded as
equal~\cite{PRL92028701,PRL93068701,PRL96218702,PRL100208701}, or
examined a single random walker travelling on weighted
networks~\cite{PRE77066105,PRE80016107}.

Before introducing our model, we firstly introduce the general
random walk on weighted networks. Let's consider a general random
walker starting from node $i$ at time step $t=0$ and denote
$P_{im}(t)$ as the probability of finding the walker at node $m$ at
time $t$. The probability of finding the walker at node $j$ at the
next time step is $P_{ij}(t+1)=\sum_{m}a_{mj}\cdot\Pi_{m\rightarrow
j}\cdot P_{im}(t)$, where $a_{mj}$ is an element of the network's
adjacent matrix. Here, $\Pi_{m\rightarrow j}$ is defined as
$\frac{w_{mj}}{s_m}$.

Thus the probability $P_{ij}(t)$ for the walker to travel from node
$i$ to node $j$ in $t$ time steps is
\begin{eqnarray}
P_{ij}(t)&=&\sum_{m_1,...,m_{t-1}}\frac{a_{im_1}w_{im_1}}{s_i}\times
\frac{a_{m_1m_2}w_{m_1m_2}}{s_{m_1}}\nonumber\\ &\times&\ldots\times
\frac{a_{m_{t-1}j}w_{m_{t-1}j}}{s_{m_{t-1}}}.
\end{eqnarray}
In other words,
$P_{ij}(t)=\sum_{m_1,...,m_{t-1}}P_{im_1}P_{m_1m_2}\cdot\cdot\cdot
P_{m_{t-1}j}$. Comparing the expressions for $P_{ij}$ and $P_{ji}$
one can see that $s_iP_{ij}(t)=s_jP_{ji}(t)$. This is a direct
consequence of the undirectedness of the network. For the stationary
solution, one obtains $P_i^\infty={s_i}/{Z}$ with $Z=\sum_is_i$.
Note the stationary distribution is, up to normalization, equal to
$s_i$, the strength of the node $i$. This means the higher strength
a node has, the more often it will be visited by a walker.

Here we propose the general random diffusion (GRD) model, which
describes the traffic fluctuation problem as a large number of
independent random walkers travelling simultaneously on a weighted
network, where a walker's choice of path is based on the rule
mentioned above.

\subsubsection{Size of time window, $M$}
We observe traffic arriving at a node (or passing through a link) in
time windows of equal size. Each time window consists of $M$ time
units, which is defined as a step for random walkers to hop from one
node to another.

\subsubsection{Preferential choice of path}
A walker at node $i$ chooses link $i$--$j$ as the next leg of travel
according to the following preferential probability,
\begin{equation}
\Pi_{i\rightarrow j}=\frac{w_{ij}}{\sum_{j\in\Gamma(i)}w_{ij}}=\frac
{w_{ij}}{s_{i}},\label{pi}
\end{equation}
which is proportional to the weight of the link.

\subsubsection{Average traffic, $\langle f\rangle$}

The traffic arriving at node $i$ during a time window is
$f_i=\sum_{m=1}^M\Delta_i(m),$ where $\Delta_i(m)$ is a random
variable representing the number of walkers arriving at node $i$ at
the $m$th time unit. The average traffic, $\langle f_i\rangle$, is
the mean traffic volume at node $i$ over all time windows.
Similarly, $f_{ij}$ is the traffic passing through a link between
nodes $i$ and $j$ during a time window, and $\langle f_{ij}\rangle$
is the average link traffic.

\subsubsection{Traffic fluctuation, $\sigma$}

The standard deviation $\sigma_i$ indicates the fluctuation of
traffic volume around the average traffic $\langle f_i\rangle$ at
node $i$ over time windows. Similarly $\sigma_{ij}$ is the
fluctuation of link traffic $f_{ij}$ on the link between nodes $i$
and $j$.

~\\

The key interest on the traffic fluctuation problem is the relation
between the average traffic $\langle f\rangle$ (of a node or link)
and the fluctuation $\sigma$, and the impact of relevant quantities
(time window size $M$, weightiness parameter $\theta$ and node
degree $k$) on such relation. In the following two sections we
investigate traffic fluctuation of node and link respectively.


\section{Node Traffic Fluctuation on Weighted Networks}\label{NTFWN}

\subsection{Analytical Solution}

According to the GRD model's preferential choice of path
(see~\ref{pi}), in the stationary regime the number of walkers visit
node~$i$ at a single time step can be estimated as
\begin{equation}
\Phi_i(r)=r\frac{s_i}{\sum_{i=1}^Ns_i},\label{Phi}
\end{equation}
where $r$ is the number of random walkers travelling on the weighted
network and $N$ is the number of nodes. In the GRD model random
walkers are independent and the arrival of walkers at a node is a
Poisson process. Thus the mean number of walkers visit node $i$ in a
window of $M$ time steps is
\begin{equation}
\langle f_i\rangle=\Phi_i(r)M, \label{fi}
\end{equation}
and the probability that exactly $n$ walkers visit node $i$ in a
time window is
\begin{equation}
P_i(n)=e^{-\Phi_i(r)M}\frac{[\Phi_i(r)M]^n}{n!}. \label{psi-simple}
\end{equation}

In a more general case, the number of walkers $r$ observed from time
window to time window is uniformly distributed in
$[R-\delta,R+\delta]$, $0<\delta\leq R$, where $R$ is the average
number of walkers and the noise constant $\delta$ is the
fluctuation. The probability of having $r$ walkers in a time window
is
\begin{equation}
F(r)=\frac{1}{2\delta+1}.\label{Fr}
\end{equation}
Then~(\ref{psi-simple}) becomes the following
\begin{eqnarray}
\psi_i(n)&=&\sum_{j=0}^{2\delta}(\frac{e^{-(s_i/\sum_{i=1}^{N}s_i)(R-\delta+j)M}}{2\delta+1}\\
\nonumber&\times&\frac{[(s_i/\sum_{i=1}^{N}s_i)(R-\delta+j)M]^n}{n!}
) . \label{psi}
\end{eqnarray}
Calculating the first and second moments of $f_i$, we get
\begin{equation}
\langle f_i\rangle=\sum_{n=0}^\infty n\psi_i(n)
=\frac{s_i}{\sum_{i=1}^Ns_i}RM \label{f_i-general}
\end{equation}
and
\begin{equation}
\langle {f^2_i}\rangle=\sum_{n=0}^\infty n^2\psi_i(n)={\langle
f_i\rangle}^2\left(1+\frac{\delta^2+\delta}{3R^2}\right)+{\langle
f_i\rangle}.
\end{equation}
Then the standard deviation as a function of $\langle f_i\rangle$ is
\begin{equation}
{\sigma^2_i}=\langle f_i\rangle\left(1+\langle
f_i\rangle\frac{\delta^2+\delta}{3R^2}\right). \label{nodesigma}
\end{equation}
This indicates the relation between the traffic at nodes and its
scale doesn't depend on the weight parameter $\theta$. The traffic
fluctuation at node $i$ can be given as $\sigma_i^2=(\sigma^{\rm
int}_i)^2+(\sigma^{\rm ext}_i)^2$. This suggests that the driving
force of traffic fluctuation at node $i$ can be ascribed to two
aspects: one is the internal randomness of the diffusion process,
$\sigma^{\rm int}_i=\sqrt{\langle f_i\rangle}$; and the other is the
change in the external environment, $\sigma^{\rm ext}_i=\langle
f_i\rangle\sqrt{\frac{\delta^2+\delta}{3R^2}}$, i.e.~the fluctuation
of the number of walkers in the network in different time windows.
To make it concrete, we will show a class of specifical networks to
testify its validity in what follows.

\subsubsection{For Neutral Weighted Networks}\label{BA}

Networks exhibit different mixing patterns, or degree-degree
correlations~\cite{PRL89208701,PRE67026126}. For example social
networks show the assortative mixing where high-degree nodes tend to
connect with other high-degree nodes and low-degree nodes with
low-degree ones. By contrast, biological and technological networks
show the disassortative mixing where high-degree nodes tend to
connect with low-degree nodes and vice versa.

Neutral networks show neither assortative nor disassortative mixing.
Three popular examples are (1)~the Erd\"{o}s-R\'{e}nyi (ER) random
graph~\cite{PM6290}, which is generated by random link attachment
between nodes and is characterised by a Poisson degree distribution;
(2)~the Barab\'asi-Albert (BA) scale-free graph~\cite{SCI286509},
which is generated by the so-called preferential attachment and is
characterised by a power-law degree distribution; and (3) the
Watts-Strogatz (WS) small-world~\cite{NATURE393440} is generated by
rewiring links on regular lattices and is characterised by a high
clustering~\cite{SIAMR45167} and a small average topological
distance~\cite{PRE70056110}. These three generic models have been
widely studied in network research.

\subsubsection{Node Strength Expressed As Node Degree}

$P(k_q|k_i)$ is the conditional probability distribution that a
$k_q$-degree node connects with a $k_i$-degree
node~\cite{PRL89208701}. For neutral networks,
\begin{equation}
P(k_q|k_i)=P(k_q|k)=k_qP(k_q)/\langle k\rangle,\label{meanfield}
\end{equation}
where $P(k_q)$ is the probability of a node having degree $k_q$ and
$\langle k\rangle$ is the average degree~\cite{PRL89208701}.

The nearest-neighbours average degree of node $i$ can be estimated
as ${{k_{nn}(k_i)}}=\sum_{k_q=k_{min}}^{k_{max}}{k_q}P(k_q|k_i),$
where suffix $q$ stands for a neighbor of node $i$, $k_{min}$ and
$k_{max}$ are the minimum and maximal node degrees in the network,
respectively. By mean-field approximation we have
\begin{equation}
{{k^\theta_{nn}(k_i)}}=\sum_{k_q=k_{min}}^{k_{max}}k_q^{\theta+1}P(k_q)/\langle
k\rangle=\langle k^{\theta+1}\rangle/\langle k\rangle. \label{knn}
\end{equation}
One can see that ${k_{nn}^\theta}(k_i)$ does not depend on the
degree $k_i$, hence
$\sum_{q\in\Gamma(i)}{k_q^{\theta}}=k_i\cdot{k^{\theta}_{nn}(k_i)}$.
Using Eqs.~(\ref{si}) and (\ref{knn}) we have
\begin{equation}
s_i={k^{\theta+1}_i}{\langle k^{\theta+1}\rangle}/{\langle
k\rangle}. \label{s(k)}
\end{equation}

For neutral weighted networks, we can then
rewrite~(\ref{f_i-general}) as
\begin{equation}
\langle f_i\rangle=\frac{k^{\theta+1}_i}{N\langle
k^{\theta+1}\rangle}RM. \label{f_i}
\end{equation}
When the four quantities $M$, $R$, $\delta$ and $\theta$ satisfy the
condition that
$\frac{\delta^2+\delta}{3R}\cdot\frac{k_i^{\theta+1}}{N\langle
k^{\theta+1}\rangle}M\ll1$, the relation between traffic fluctuation
and average traffic as given in~(\ref{nodesigma}) is reduced to a
power-law scaling $\sigma\sim{\langle f\rangle}^{\alpha}$ with
$\alpha=1/2$. When
$\frac{\delta^2+\delta}{3R}\cdot\frac{k_i^{\theta+1}}{N\langle
k^{\theta+1}\rangle}M$ is not negligible, the exponent $\alpha$ is
in the range of $[1/2, 1]$.

\subsection{Numerical Simulation}

We run numerical simulations for the following purpose: (1) to
verify the analytical solution; (2) to examine the impact of
parameters such as window size $M$ and node degree $k$ on the
power-law scaling of the traffic fluctuation function; and (3) to
contrast unweighted networks with $\theta=0$ against weighted
networks with $\theta>0$.

\subsubsection{Simulation Settings}

Our simulation is based on network graphs generated by the BA
model~\cite{SCI286509}, which is neutral mixing and features a
power-law degree distribution $P(k)\sim k^{-3}$.
We generate ten BA graphs, each of which has 5,000 nodes and 25,000
links. We assign link weight and node strength as defined
in~(\ref{wij}) and~(\ref{si}) respectively.
Initially, we disperse $r=R\pm\delta=10,000\pm1,000$ random walkers
uniformly on nodes.
At each time step, all walkers travel one hop according
to~(\ref{pi}).
For a given time window size of $M$, we observe traffic fluctuation
at each node over a large number of time windows.

For each given value of time window size $M$ or weightiness
parameter $\theta$, we repeat the simulation for 50 times (with
different random seeds) on each of the ten BA networks. Each result
shown below is averaged over the $10\times50=500$ simulations.

\subsubsection{Power-Law Relation Between $\sigma_i$ and $\langle
f_i\rangle$}

\begin{figure}
\begin{center}
\includegraphics[width=8cm]{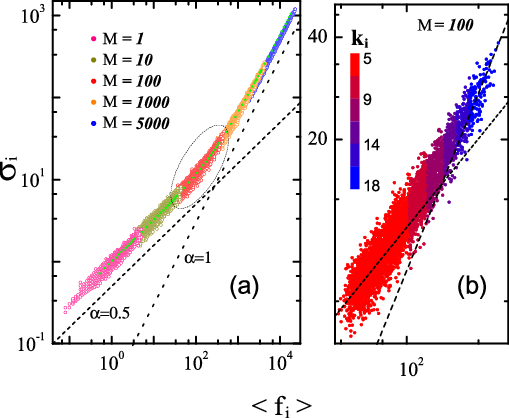}
\caption{Traffic fluctuation $\sigma_i$ as a function of average
traffic $\langle f_i\rangle$ at node $i$. ({\bf a}) Observed in
different time window sizes of $M=1, 10, 100, 1000$ and 5000; and
({\bf b}) for nodes of different degrees (with $M=100$). The green
dotted-line in ({\bf a}) is the analytical solution given
in~(\ref{nodesigma}). The two black dotted-lines in both ({\bf a})
and ({\bf b}) correspond to $\sigma_i\sim \langle
f_i\rangle^{\alpha}$ with $\alpha=0.5$ and 1, respectively.
Simulation results are obtained on weighted BA networks having 5,000
nodes and 25,000 links with the weightiness parameter $\theta=0.5$,
the average number of random walkers $R=10^4$ and the noise constant
of $\delta=10^3$. For clarity we only show nodes with degrees
smaller than 18. Note that the minimal degree in the networks is 5
as $m_0=5$ and $m=5$ in the weighted BA networks.}\label{fig1}
\end{center}
\end{figure}

Figure~\ref{fig1}(a) shows the relation between traffic fluctuation
$\sigma$ and average traffic $\langle f\rangle$ for different time
window size $M$ where the weightiness parameter is set as
$\theta=0.5$. The simulation results overlap with the analytical
solution. Both the average traffic and the traffic fluctuation
increase with the size of window $M$. For any given value of $M$,
the two quantities follow a power-law relation $\sigma\sim \langle
f\rangle^{\alpha}$. When $M$ is small, the power-law exponent
$\alpha$ is close to $1/2$; and when $M$ increases the exponent
grows towards $1$.

Figure~\ref{fig1}(b) shows the enlargement of the traffic
fluctuation function for the window size $M=100$ as circled out in
Figure~\ref{fig1}(a), where data dots are coloured by node degrees.
For nodes with the higher degree, the large values of $\sigma$ and
$\langle f\rangle$ are observed. For low-degree nodes (e.g. $k=5$)
the power-law exponent is close to $\alpha=1/2$; whereas for higher
degree nodes (e.g. $k=18$), the exponent approaches to $1$.

As predicted by Eq.~(\ref{f_i}), our simulation results confirm that
the traffic fluctuation function $\sigma\sim{\langle
f\rangle}^{\alpha}$ does not follow a simple power-law. Rather, the
power-law scaling is in the range of [1/2, 1]. It is affected by a
number of parameters including the window size $M$ and the degree of
nodes under study. This echoes previous studies as their random
diffusion model based on unweighted networks is a special case of
our general random diffusion model on weighted networks.

\subsubsection{Impact of Weightiness Parameter $\theta$}

\begin{figure*}
\begin{center}
\includegraphics[width=0.45\linewidth]{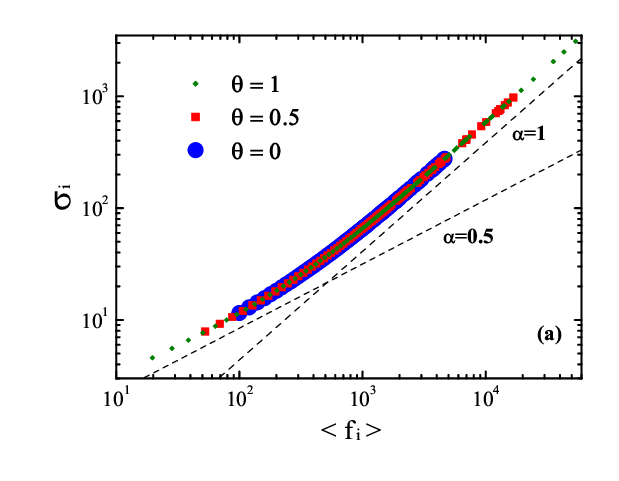}\hspace{1cm}
\includegraphics[width=0.45\linewidth]{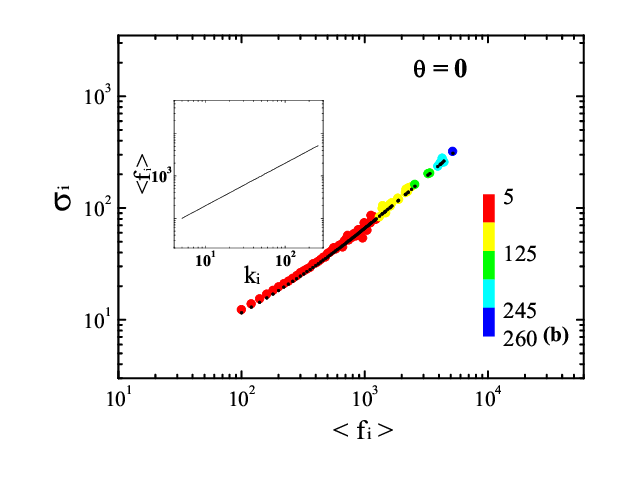}
\includegraphics[width=0.45\linewidth]{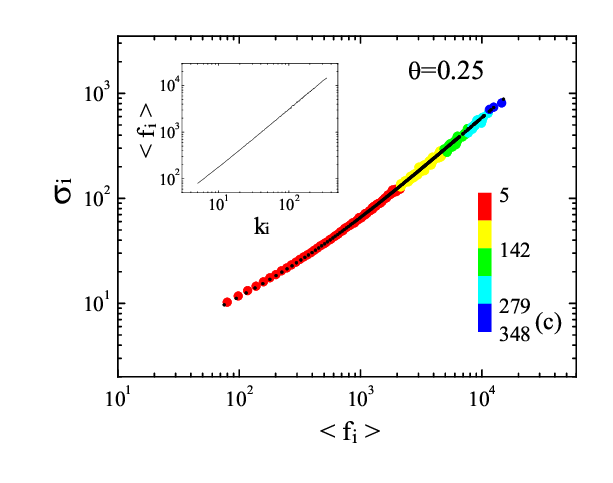}\hspace{1cm}
\includegraphics[width=0.45\linewidth]{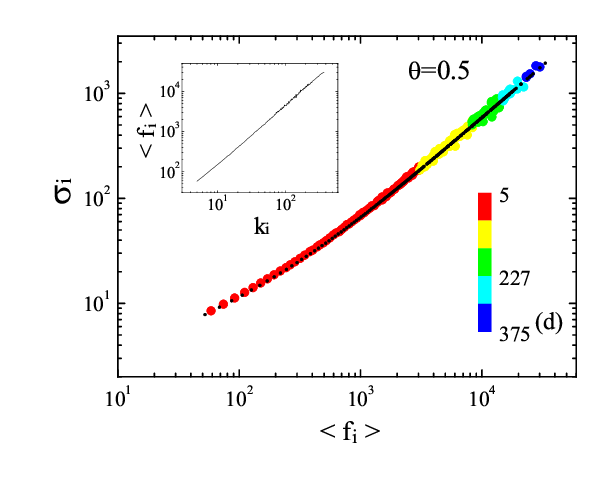}
\includegraphics[width=0.45\linewidth]{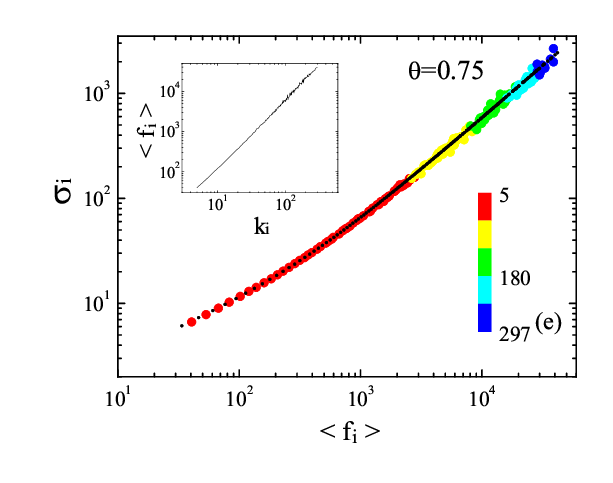}\hspace{1cm}
\includegraphics[width=0.45\linewidth]{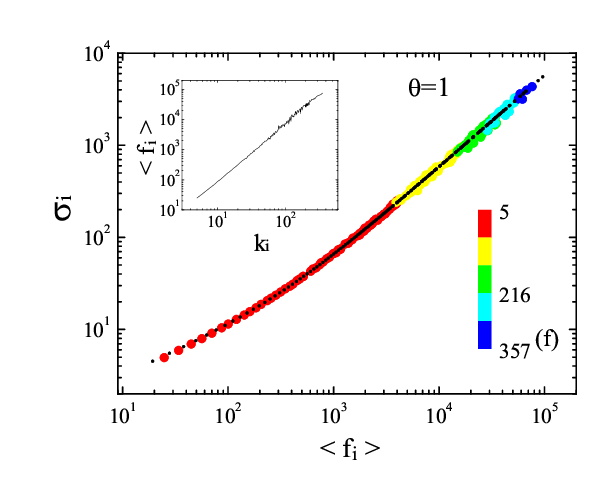}
\caption{Traffic fluctuation $\sigma_i$ as a function of average
traffic volume $\langle f_i\rangle$ at node $i$ for different values
of weightiness parameter $\theta$. ({\bf a}) shows analytical
solutions of~(\ref{nodesigma}) for $\theta=0$, 0.5, and~1; and ({\bf
b}), ({\bf c}), ({\bf d}), ({\bf e}) and ({\bf f}) show simulation
results for $\theta=0$, 0.25, 0.5, 0.75, and~1 respectively, where
results are shown for all nodes and coloured by node degree $k_i$,
and the black dotted lines are the analytical solutions. Time window
size $M$ is set as 100 and other parameters are as before. The
insets of ({\bf b}), ({\bf c}) and ({\bf d}) show $\langle
f_i\rangle$ as a function of $k_i$ on a log-log scale, which is
predicted by~(\ref{f_i}).} \label{fig2}
\end{center}
\end{figure*}

\begin{figure}
\begin{center}
\scalebox{0.8}[1]{\includegraphics[trim=30 10 0 0]
{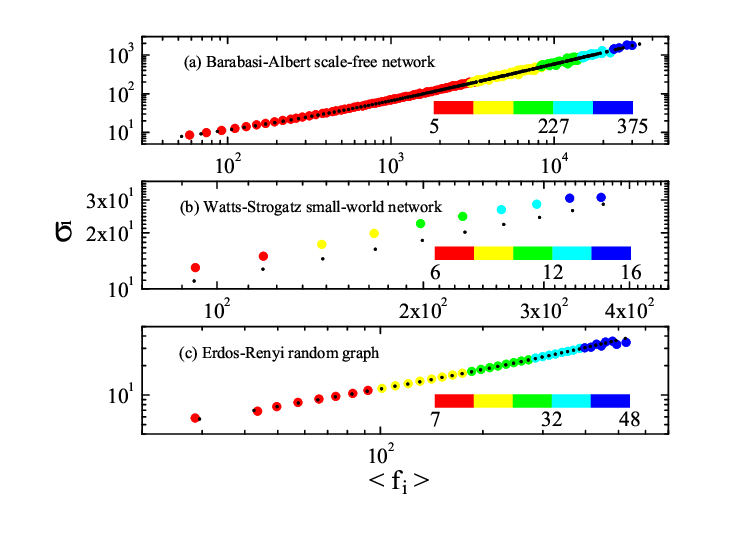}}
\caption{Traffic fluctuation $\sigma_{i}$ as a function of average
traffic volume $\langle f_i\rangle$ at node $i$ for neutral networks
with $\theta=0.5$, $M=100$, $R=10^4$ and $\delta=10^3$.
The black dotted lines are the analytical solution given
by~(\ref{nodesigma}) and~(\ref{f_i}).
({\bf a}) shows the simulation obtained on weighted BA networks with
5,000 nodes and 25,000 links.
({\bf b}) shows the simulation obtained on weighted Watts-Strogatz
small-world networks with rewiring probability 0.1, 5,000 nodes
and 25,000 links.
({\bf c}) shows the simulation obtained on weighted
Erd\"{o}s-R\'{e}nyi random graphs with connected probability 0.002,
5,000 nodes and 25,000 links.}\label{Fig3_F_K_COM}
\end{center}
\end{figure}

Figure~\ref{fig2}(a) illustrates the solutions of~(\ref{nodesigma})
for the weightiness parameter $\theta=0$, 0.5 and 1 with the time
window size is set as $M=100$. Figure~\ref{fig2}(b), (c), (d), (e)
and (f) show the simulation results for $\theta=0$, 0.25, 0.5, 0.75,
and~1 respectively.
For different $\theta$ values, the traffic fluctuation curves
overlap with each other, and in all cases the high-degree nodes are
concentrated at the upper-right end of the curves whereas the
low-degree nodes are dispersed alone the lower-left part of the
curve.
The remarkable difference, however, is that with the increase of
$\theta$ the value ranges of $\langle f_i\rangle$ and $\sigma_i$
expand significantly towards both directions.
This means that comparing with an unweighted network, traffic
fluctuation in a weighted network is more acute at high-degree nodes
and more stable at low-degree nodes. This is because in a weighted
network the node strength $s\sim k^{\theta+1}$ (see~(\ref{s(k)}))
and therefore high-degree nodes deprive more traffic from low-degree
ones than in an unweighted network.

\subsubsection{For Neutral Weighted Networks}\label{Deviation}

Figure~\ref{Fig3_F_K_COM} shows the simulation results on weighted
BA, Watts-Strogatz small-world~\cite{NATURE393440} networks and
Erd\"{o}s-R\'{e}nyi random graphs with $\theta=0.5$. As shown in the
panels (a) and (c), the simulation results on the scale of $f_i$ is
remarkably consistent with the solution given by
Eq.~\ref{nodesigma}. However, one can find that the numerical
results in the panel (b) deviate from Eq.~\ref{nodesigma}
apparently. This deviation is due to most of the links (at least
$80\%$ in our simulations) in the small-world networks are still
regularly connected. $80\%$ is predicted by the small rewiring
probability $0.1$. Obviously, the conditional probability distribution
$P(k_q|k_i)$ doesn't completely match Eq.~\ref{meanfield}.

Our results suggest that if a real system should be described as a
weighted network with $\theta=1$ but instead an unweighted network
with $\theta=0$ is used, then we would underestimate the $\langle
f_i\rangle$ and $\sigma_i$ values for high-degree nodes and
overestimates the values for low-degree nodes by as large as one
order of magnitude.
This highlights the importance of choosing a proper network model
for traffic fluctuation research.

\section{Link Traffic Fluctuation on Weighted Networks}

\subsection{Analytical Solution}

In GRD model, random walkers on a weighted network travel
independently and therefore the number of walkers passing through a
link is a Poisson process. As given in~(\ref{pi}), the probability
that a walker at node $i$ chooses link $i$--$j$ as the next leg of
travel is ${w_{ij}}/s_{i}$. Thus for $r$ random walkers in a
weighted network, the average number of walkers passing through link
$i$--$j$ (from node $i$ to node $j$ as well as from node $j$ to node
$i$) during a time window $M$ is
$$\langle f_{ij}\rangle=\Omega_{ij}(r)M$$ where
\begin{equation}
\Omega_{ij}(r)=r\left(\frac{s_i}{\sum_{i=1}^{N}s_i}\cdot\frac{w_{ij}}{s_i}+\frac{s_j}{\sum_{i=1}^{N}s_i}\cdot\frac{w_{ij}}{s_j}\right),
\end{equation}
and the probability of $f_{ij}=n$ in a time window is
\begin{equation}
Q_{ij}(n)=e^{-\Omega_{ij}(r)M}\frac{[\Omega_{ij}(r)M]^n}{n!}.
\end{equation}

Similar as the above analysis on node traffic fluctuation, for a
more general case where the number of random walkers $r$ from time
window to time window is distributed in $[R-\delta,R+\delta]$, the
probability of $f_{ij}=n$ in a time window is
\begin{eqnarray}
\Gamma_{ij}(n)&=&\sum_{j=0}^{2\delta}(\frac{e^{\frac{2w_{ij}}{\langle
k^{\theta+1}\rangle^2N}(R-\delta+j)M}}{2\delta+1}\\\nonumber&\times&\frac{e^{\frac{2w_{ij}}{{\langle
k^{\theta+1}\rangle}^2N}(R-\delta+j)M}}{n!}).
\end{eqnarray}
Calculating the first and second moments of $f_{ij}$, we obtain
\begin{equation}
\langle f_{ij}\rangle=\sum_{n=0}^\infty n\Gamma_{ij}(n)
=\frac{2w_{ij}}{\sum_{i=1}^Ns_i}RM, \label{fij-general}
\end{equation}
and
\begin{equation}
\langle f^2_{ij}\rangle=\sum_{n=0}^\infty n^2\Gamma_{ij}(n)={\langle
f_{ij}\rangle}^2\left(1+\frac{\delta^2+\delta}{3R^2}\right)+{\langle
f_{ij}\rangle}.
\end{equation}
Thus the standard deviation as a function of the average traffic
$\langle f_{ij}\rangle$ is
\begin{equation}
{\sigma^2_{ij}}=\langle f_{ij}\rangle\left(1+\langle
f_{ij}\rangle\frac{\delta^2+\delta}{3R^2}\right). \label{fijsigma}
\end{equation}
This indicates the relation between the traffic on links and its
scale is irrelevant to $\theta$ as well. For neutral weighted
networks, using Eqs.~(\ref{wij}) and (\ref{s(k)}), we can
rewrite~(\ref{fij-general}) as
\begin{equation}
\langle f_{ij}\rangle =\frac{2(k_ik_j)^\theta\langle k\rangle
MR}{{\langle k^{\theta+1}\rangle}^2N}.\label{f_ij}
\end{equation}
If $\frac{\delta^2+\delta}{3R}\cdot\frac{2(k_ik_j)^\theta\langle
k\rangle M}{{\langle k^{\theta+1}\rangle}^2N}\ll1$,~(\ref{fijsigma})
is reduced to a power-law scaling $\sigma_{ij}\sim{\langle
f_{ij}\rangle}^{\alpha}$ with $\alpha=1/2$. Conversely, as
$\frac{\delta^2+\delta}{3R}\cdot\frac{2(k_ik_j)^\theta\langle
k\rangle M}{{\langle k^{\theta+1}\rangle}^2N}$ increases to $1$, the
exponent $\alpha$ will leave $1/2$ for $1$.

\subsection{Numerical Simulation}

Here we use the same simulation settings as Section~\ref{NTFWN}(B).

\begin{figure}
\begin{center}
\includegraphics[width=8.5cm]{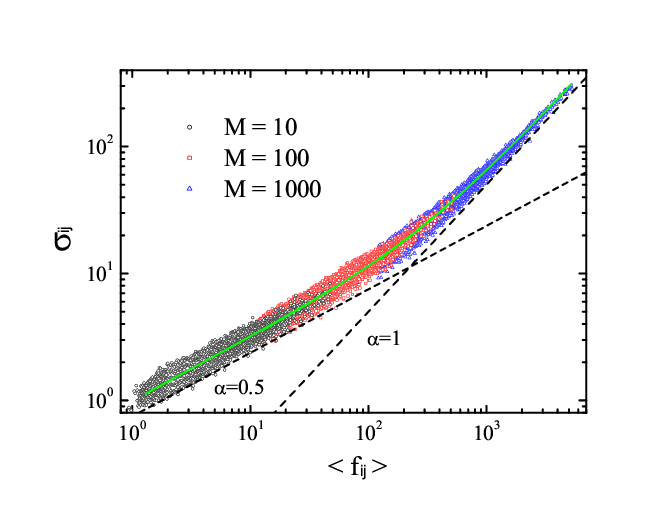}
\caption{Traffic fluctuation $\sigma_{ij}$ as a function of average
traffic $\langle f_{ij}\rangle$ on link $i$--$j$ between nodes $i$
and $j$ with time window size of $M=10$, 100 and 1000.
The green line is the analytical solution given by~(\ref{fijsigma}).
The black dotted lines correspond to $\sigma_{ij}\sim \langle
f_{ij}\rangle^{\alpha}$ with $\alpha=0.5$ and $1$, respectively.
The simulation results are obtained on weighted BA networks with
5,000 nodes and 25,000 links with $\theta=0.5$, $R=10^4$ and
$\delta=10^3$. }\label{fig4_M}
\end{center}
\end{figure}

\subsubsection{Power-Law Relation Between $\sigma_{ij}$ and $\langle
f_{ij}\rangle$}

In Figure~\ref{fig4_M} we plot the relation between the traffic
fluctuation $\sigma_{ij}$ and the average traffic $\langle
f_{ij}\rangle$ on link $i$--$j$ for three different values of time
window size $M$. The simulation results are in agreement with our
analytical solution. As predicted by Eqs.~(\ref{fijsigma}) and
(\ref{f_ij}), the average traffic and the fluctuation increase with
$M$. The two quantities follow a power-law scaling
$\sigma_{ij}\sim{\langle f_{ij}\rangle}^{\alpha}$, where the
exponent $\alpha$ is $1/2$ for small value of $M$ and approaches to
1 with larger $M$. Such behaviour is similar as the traffic
fluctuation on nodes.

\subsubsection{Impact of Weightiness Parameter $\theta$}

\begin{figure*}
\begin{center}
\includegraphics[width=0.45\linewidth,trim=0 1 0 0]{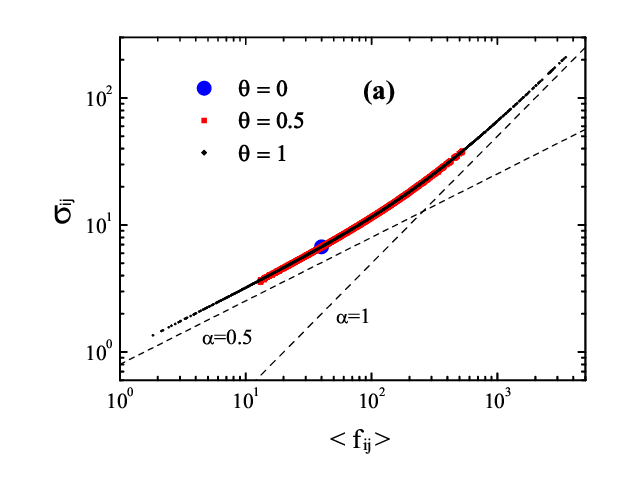}\hspace{1cm}
\includegraphics[width=0.45\linewidth]{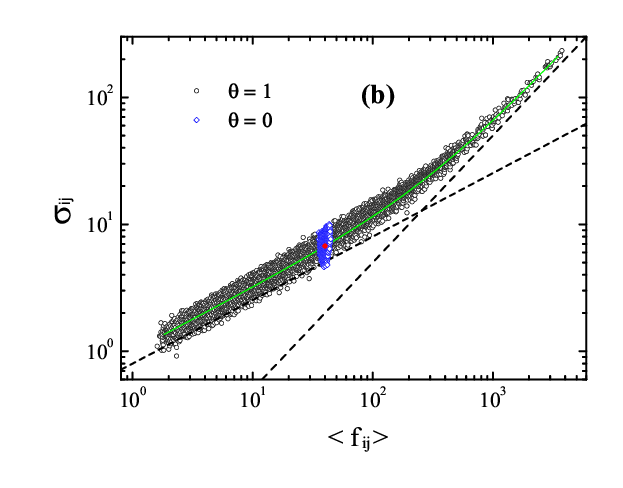}
\includegraphics[width=0.45\linewidth,trim=0 1 0 0]{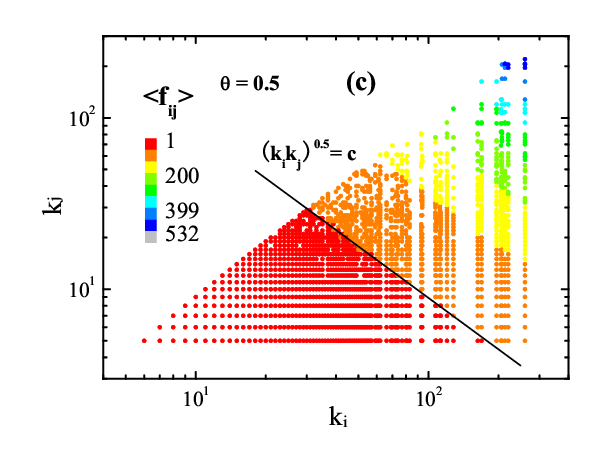}\hspace{1cm}
\includegraphics[width=0.45\linewidth]{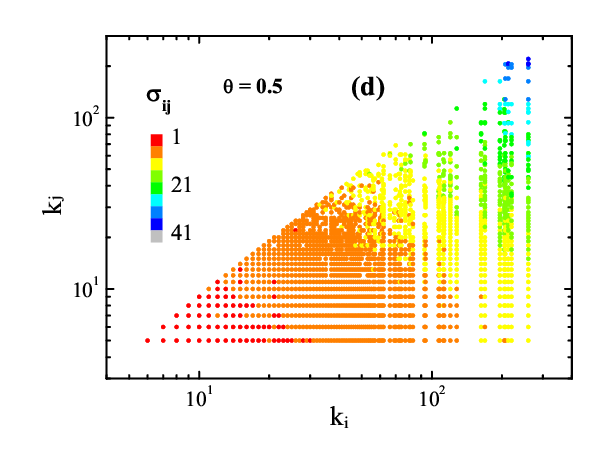}
\caption{ Traffic fluctuation $\sigma_{ij}$ and average traffic
$\langle f_{ij}\rangle$ on link $i$--$j$ with parameters $M=100$,
$R=10^4$ and $\delta=10^3$.
({\bf a}) Shows analytical solutions of~(\ref{fijsigma}) for
$\theta=0$, 0.5 and~1, respectively.
({\bf b}), ({\bf c}) and ({\bf d}) are simulation results. ({\bf b})
shows $\sigma_{ij}$ as a function of $\langle f_{ij}\rangle$ for
$\theta=0$ and 1.
({\bf c}) shows the average traffic $\langle f_{ij}\rangle$ on link
$i$--$j$ as a function of degrees of the two end nodes of the link,
$k_i$ and $k_j$, where $\theta=0.5$, the value of $\langle
f_{ij}\rangle$ is given by a colour bar, and the guideline is given
by~(\ref{f_ij}). Similarly ({\bf d}) shows the traffic fluctuation
$\sigma_{ij}$ as a function of $k_i$ and $k_j$. }\label{fig5ij}
\end{center}
\end{figure*}
\begin{figure}
\begin{center}
\scalebox{0.8}[1]{\includegraphics[trim=20 10 0
0]{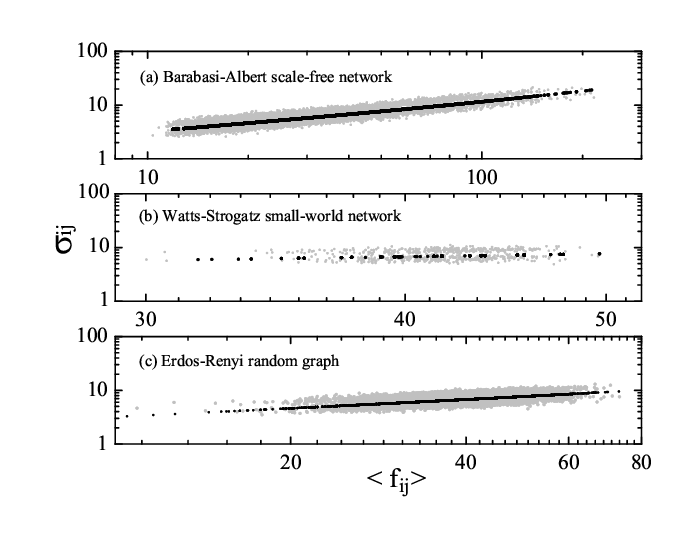}}
\caption{Traffic fluctuation $\sigma_{ij}$ and average traffic
$\langle f_{ij}\rangle$ on link $i$--$j$ with parameters $M=100$,
$R=10^4$ and $\delta=10^3$.
The black dotted lines are the analytical solution given
by~(\ref{fijsigma}) and~(\ref{f_ij}).
({\bf a}) shows the simulation obtained on weighted BA networks with
5,000 nodes and 25,000 links with $\theta=0.5$, $R=10^4$ and
$\delta=10^3$.
({\bf b}) shows the simulation obtained on weighted Watts-Strogatz
small-world networks with a rewiring probability 0.1, 5,000 nodes
and 25,000 links.
({\bf c}) shows the simulation obtained on weighted
Erd\"{o}s-R\'{e}nyi random graphs with connected probability 0.002,
5,000 nodes and 25,000 links.}\label{Fig6_F_EDGE_COM}
\end{center}
\end{figure}

In Figure~\ref{fig5ij}(a), the range of scale for $\theta=1$ is
$[0.5031, 0.9602]$ while for $\theta=0$ it is nearly $0$. For
$\theta=0$, the links in the simulation form a dense group on the
plot~\cite{NJP9154}, representing almost equal fluctuation
properties. This unaccounted fact can be explained by the solutions
of~(\ref{fijsigma}) and~(\ref{f_ij}). The dashed lines are guides to
the eyes and correspond to $\sigma_i\sim {\langle
f_i\rangle}^\alpha$, with $\alpha=1/2$ and $\alpha=1$. The
comparison among unweighted networks ($\theta=0$) and weighted ones
($\theta>0$) can be observed in this figure in panel (b). Note that
the green and red dots reflect the solution of~(\ref{fijsigma}). As
shown in the figure, the differences of $f_{ij}$ and $\sigma_{ij}$
for different nodes pairs crop up when $\theta=1$. In fact, this
process is not as sudden as it looks.

\subsubsection{Node Degree $k$}

In Figure~\ref{fig5ij}(c) and (d), we show the middle case of
$\theta=0.5$ numerically for different node pairs. As shown in the
panel (c), the plots are colored by $f_{ij}$. For all the links, we
only focus on the results obtained for pairs of $k_i$ and $k_j$
(restrict $k_j>10$ to enhance the speed of loading figures), where
$k_i>k_j$. One can easily find that $f_{ij}$ is directly
proportional to the product of two ends' degrees $k_i$ and $k_j$
when $\theta>0$, while they are almost a constant when $\theta=0$.
Likewise, $\sigma_{ij}$'s are directly proportional to $k_ik_j$ when
$\theta=0.5$ as well, but they are rather stable when $\theta=0$
(see panel (b)).

\subsubsection{For Neutral Weighted Networks}

Figure~\ref{Fig6_F_EDGE_COM} shows the simulation results on
weighted BA, Watts-Strogatz small-world networks and
Erd\"{o}s-R\'{e}nyi random graphs with $\theta=0.5$. As shown in the
panels (a) and (c), the simulation results on the scale of $f_{ij}$
is consistent with the solution given by Eq.~\ref{fijsigma}. In the
panel (b), one can find that the numerical results deviate from
Eq.~\ref{fijsigma} again. The behavior confirms our observations in
Fig.~\ref{Fig6_F_EDGE_COM} and the discussion in
Section~\ref{Deviation} in another light.

\subsection{Discussion}

One simple example for the result is that for a traffic network, the
traffics on different roads differ, the wider of which can have the
larger traffic. At the same time, the roads with heavy loads
fluctuate more dramatically, depending on whether it is a rush hour
or not. Indeed, the interactions among walkers should not be ignored
in the realistic scenarios, e.g., the transport of information
packets, signals, molecules, rumours, diseases, to name but a few.
Whereas, these interactions vary widely from case to case. For
generality, we begin with this simple model to take a step forward
in the analytical investigation of the corresponding problems. We
believe our rigorous solutions are capable of prompting related
studies on interacting walkers in the near future.

\section{\label{sec:conclusion}CONCLUSION\protect}

In summary, we investigate the traffic fluctuation problem on
weighted networks, which is a more general and realistic
representation of real systems. Previous results on nodes are the
most simple case of our result. Moreover, comparatively few
investigations have been recorded in the literature relative to the
fluctuation on links both for unweighted and weighted networks yet.

In this paper, we introduce the general random diffusion (GRD)
model, which describe a swarm of random walkers traveling
simultaneously on the weighted networks. Based on the model we
provide analytic solutions to characterise the relation between the
mean traffic and its fluctuation for nodes and for links. We discuss
the impact of key parameters on the traffic fluctuation. Key
observations include size of time window, node strength, and link
weight. To prove the results, we take neutral networks with a
specifical link weight definition for example. Our analysis
indicates the relation between the traffic and its scale is
irrelevant to the weight parameter $\theta$. Simultaneously, we find the scales
of traffic on weighted links with $\theta>0$ are much wider than
unweighted ones, on which the traffic are rather stable, which is
confirmed by analytical prediction with remarkable accuracy. Thus,
both simulations and analytic work have suggested that the weight
could have an impact on the way in which networks operate, including
the way information travels through the network and resource
assignment for an efficient performance of communication networks.

One significant observation is that study based on unweighted
networks could lead to unrealistic, misleading results, such as same
link traffic and fluctuation for all links. Therefore, the GRD model
based on weighted network provides a proper platform for
future such research.


\begin{acknowledgments}
We thank Ming Tang, Wenxu Wang, Xiangwei Chu and Xiaoming Liang for
useful discussions. This research was supported by the National
Natural Science Foundation of China under Grant Nos. 860873040 and
60873070, and 863 program under Grant No. 2009AA01Z135. Jihong Guan
was also supported by the ``Shuguang Scholar" Program of Shanghai
Education Development Foundation under grant No. 09SG23. Shi Zhou is
supported by the Royal Academy of Engineering and the Engineering
and Physical Sciences Research Council (UK) under grant
no.~10216/70.
\end{acknowledgments}


\end{document}